\documentclass[aps,prd,floatfixy,superscriptaddress,showpacs,showkeys]{revtex4}[14pt]
\usepackage{amssymb}
\usepackage{amsmath}
\usepackage{amsfonts}
\usepackage{epsfig}
\usepackage{graphicx}
\usepackage{tabularx}
\usepackage{latexsym}
\usepackage{subfigure}
\usepackage{verbatim}
\usepackage{color}
\usepackage{braket}
\usepackage{multirow}
\usepackage{dcolumn}
\usepackage{bm}
\newcommand{\seq}{\begin{subequations}}
\newcommand{\sen}{\end{subequations}}
\newcommand{\eq}{\begin{eqnarray}}
\newcommand{\en}{\end{eqnarray}}

\def\ds{D^{\ast  0}}

\def\L2{\Lambda^2}

\def\eq{\begin{eqnarray}}
\def\en{\end{eqnarray}}

\usepackage[normalem]{ulem}

\renewcommand\sout{\bgroup \color{red} \ULdepth=-.5ex \ULset}

\def\ds{D^{\ast  0}}

\def\L2{\Lambda^2}

\def\eq{\begin{eqnarray}}
\def\en{\end{eqnarray}}

\def\L{{\cal L}}

\def\ds{d^*}
\def\vk{\vec{k}}

\begin{document}

\title{\bf \boldmath Decay width of $d^*(2380)
\to NN \pi$ process in a chiral constituent quark model}
\vspace{1.5cm}
\author{Yubing Dong}
\affiliation{Institute of High Energy Physics, Chinese Academy of
Sciences, Beijing 100049, China}
\affiliation{Theoretical Physics
Center for Science Facilities (TPCSF), CAS, Beijing 100049, China}
\affiliation{School of Physical Sciences, University of Chinese
Academy of Sciences, Beijing 101408, China}

\author{Fei Huang}
\affiliation{School of Physical Sciences, University of Chinese
Academy of Sciences, Beijing 101408, China}

\author{Pengnian Shen}
\affiliation{College of Physics and Technology, Guangxi Normal
University, Guilin  541004, China}
\affiliation{Institute of High
Energy Physics, Chinese Academy of Sciences, Beijing 100049, China}
\affiliation{Theoretical Physics Center for Science Facilities
(TPCSF), CAS, Beijing 100049, China}

\author{Zongye Zhang}
\affiliation{Institute of High Energy Physics, Chinese Academy of
Sciences, Beijing 100049, China} \affiliation{Theoretical Physics
Center for Science Facilities (TPCSF), CAS, Beijing 100049, China}
\affiliation{School of Physical Sciences, University of Chinese
Academy of Sciences, Beijing 101408, China}

\date{\today}

\begin{abstract}
The width of three-body single-pion decay process $d^*\to
NN\pi^{0,\pm}$ is calculated by using the $d^*$ wave function
obtained from our chiral SU(3) constituent quark model calculation.
The effect of the dynamical structure on the width of
$d^*$ is taken into account in both the single $\Delta\Delta$
channel and coupled $\Delta\Delta+CC$ two-channel
approximations. Our numerical result shows that in the
coupled-channel approximation, namely, the hidden-color
configuration being considered, the obtained partial decay width of
$d^*\to NN\pi$ is about several hundred $\rm {KeV}$,
while in the single $\Delta\Delta$ channel
it is just about $2\sim 3~\rm{MeV}$. We, therefore, conclude
that the partial width in the single-pion decay process of
$d^*$ is much smaller than the widths in
its double-pion decay processes. Our prediction may
provide a criterion for judging different
interpretations of the $d^*$ structure, as different pictures
for the $d^*$ may result quite different partial decay width.
\end{abstract}



\maketitle


In recent years, CELSIUS/WASA and WASA@COSY Collaborations~\cite{CELSIUS-WASA,CELSIUS-WASA1}
have reported a clear evidence of a resonance-like structure in double
pionic fusion channels $pn\to d\pi^0\pi^0$ and $pn\to d\pi^+\pi^-$
when they studied the ABC effect\cite{ABC} and when they treated the
neutron-proton scattering data with newly measured analyzing power
$A_y$. Since the observed structure cannot be simply understood by either the intermediate
Roper excitation contribution or the t-channel $\Delta\Delta$
process, they proposed an assumption of existing a
$d^*$ resonance whose quantum number, mass, and
width are $I(J^P)=0(3^+)$, $M \approx 2370$ MeV and
$\Gamma \approx 70$ MeV~\cite{CELSIUS-WASA,CELSIUS-WASA1} (see also
their recent paper~\cite{Bashkanov}, the averaged mass and width are
$M \approx 2375$ MeV and $\Gamma \approx 75$ MeV, respectively). Due
to its baryon number being 2, it would be treated as a dibaryon, and
could be explained by either "an exotic compact particle or a
hadronic molecule"~\cite{CERN}. Moreover, according to the
experimental data, the mass of the $d^*$ is about 80~MeV below the
$\Delta\Delta$ threshold and about $70~MeV$ above the $\Delta\pi
N$ threshold, therefore, the threshold (or cusp) effect is
expected not to be so significant as that in the XYZ study (see the
review of XYZ particles ~\cite{Chen:2016qju}).Thus, understanding
the internal structure of $d^*$ would be of great interest.\\

Actually, the existence of such a non-trivial six-quark
configuration with $I(J^P)=0(3^+)$ (called $d^*$ lately) has
triggered a great attention and has intensively been studied in the
literature since Dyson's pioneer
work~\cite{Dyson,Thomas,Oka,Wang,Yuan,Kukulin}. It should specially
be mentioned that one of those calculations reported in 1999 studied
the binding behavior of the $3^+$ dibaryon system by taking into
account a $\Delta\Delta$ channel and a hidden-color channel (denoted
by $CC$ hereafter) simultaneously~\cite{Yuan}. In that paper, the
binding energy was predicted to be about $40-80$ MeV which is quite
close to the recent observation, and the importance of the $CC$
channel was particularly emphasized. Unfortunately, the
width of the state was not calculated.\\

After the discovery of $\ds$, there are mainly three types of
explanation for its nature. Based on the SU(2) quark model,
Ref. \cite{Wang1} proposed a $\Delta\Delta$ resonance structure and
performed a multi-channel scattering calculation. They obtained a
binding energy of about 71 MeV (namely $M_{d^*}=2393$ MeV) and a
width of about $150$ MeV which is apparently much larger than the
observation. On the other hand, Ref.~\cite{Gal} studied a three-body
system of $\Delta N \pi$ and found a resonance pole with a mass of
$2363\pm 20$ MeV and a width of $65\pm 17$ MeV. An important view
point, claimed by Bashkanov, Brodsky and Clement \cite{Brodsky} in
2013, is that a dominant hidden-color structure (or six-quark
configuration) of $d^*$ is necessary for understanding the compact
structure of $d^*$. Sooner after, following our previous
prediction~\cite{Yuan}, Huang and his collaborators made an explicit
dynamical calculation by using a chiral SU(3) constituent quark
model~\cite{Zhang1,Zhang2,Huang2} in the framework of the Resonating
Group Method (RGM), and showed that the $\ds$ state has a mass of
$2380-2414$ MeV, which agrees with COSY's observation, and does have
an explicit  ``CC" configuration of about $66-68$\% in its wave
function~\cite{Huang}. Based on the obtained wave functions of $\ds$
and deuteron, Dong and his collaborators calculated the partial
decay widths of the "Golden" decay channel $d^*\to
d+2\pi^0(\pi^+\pi^-)$~\cite{Dong} as well as the widths of its
four-body decay $\ds\to pn\pi^0\pi^0$ and $\ds\to
pn\pi^+\pi^-$~\cite{Dong1}. The results of the two papers showed that inclusion
of the $CC$ configuration inside $d^*$ would make the calculated widths
suppressed greatly and the resultant partial widths for
all the $d^*\to d\pi^0\pi^0$, $d^*\to d\pi^+\pi^-$, $\ds\to
pn\pi^0\pi^0$, and $\ds\to pn\pi^+\pi^-$ decay processes are well
consistent with the experimental data. As a consequence,
the total width of $\ds$ is about $72$ MeV,
which is fairly close to the observation. All these outcomes support
that $\ds$ is probably a six-quark dominated exotic state due to its
large CC component. The general review on the dibaryon studies can be
found in Ref.~\cite{Clement} by Clement.\\

Recently, the questions about how large the decay width of
the single pion decay mode of $\ds$ is and whether such a decay
process can be observed have been discussed. Up to now, the $\ds \to
NN\pi$ decay process has not been found in the data
analysis~\cite{Clement1}, but a sizable cross section of this
process was predicted by using a $\Delta N \pi$ model where the {\it
{rms}} of $\ds$ is about $1.5fm$~\cite{Gal1}. This contradictory
information encourages us to calculate the partial decay width of
this process in the same way employed in calculating partial decay
widths of the $d\pi\pi$ and $NN\pi\pi$ processes and with the same
$\ds$ wave function obtained in our $\Delta\Delta + CC$ model, with
which the resultant {\it {rms}} of $\ds$ is only about
$0.8fm$~\cite{Zhang1,Zhang2,Huang2}, and all the calculated partial
decay widths of the $d\pi\pi$ and $NN\pi\pi$ processes and the total
width of $\ds$ are consistent with the data quite
well~\cite{Dong,Dong1}. Therefore, the obtained partial width might
be used to distinguish the models for the structure of $\ds$.\\

Similar to our previous work~\cite{Dong,Dong1}, we employ the
phenomenological effective Hamiltonian for the pseudo-scalar
interaction among quark, pion, and quark in the non-relativistic
approximation \eq {\cal H}_{qq\pi}=
\frac{g_{qq\pi}}{(2\pi)^{3/2}\sqrt{2\omega_{\pi}}} \,\,
\vec{\sigma}\cdot\vk_{\pi}\tau\cdot\phi, \en where $g_{qq\pi}$ is
the coupling constant, $\phi$ stands for the $\pi$ meson field,
$\omega_{\pi}$ and $\vk_{\pi}$ are the energy and three-momentum of
the $\pi$ meson, respectively, and ${\bf \sigma} ({\bf \tau})$
represents the spin (isospin) operator of a single quark. The wave
functions of the nucleon and $\Delta(1232)$ resonance in the
conventional constituent quark model can be found in \cite{Dong}.
The experimental data for the decay width of $\Delta\to \pi N$ is
$\sim 117\rm{MeV}$, and the theoretical calculation gives
$\Gamma_{\Delta\to\pi
N}=\frac{4}{3\pi}k_{\pi}^3(g_{qq\pi}I_o)^2E_N/M_{\Delta}$, where
$E_{N}=\sqrt{M_{N}^2+\vec{p}_N^2}$ is the energy of the outgoing
nucleon,  $k_{\pi}\sim 0.229~\rm{GeV}$ is the three-momenta of pion,
and $I_o$ denotes the spatial overlap integral of the internal wave
functions of the nucleon and the $\Delta$ resonance,
we can extract the coupling constant $g_{qq\pi}$
(the details can be found in Ref. \cite{Dong}). Defining
$G=g_{qq\pi}I_o$, the obtained $G$ value is about $5.41$ GeV$^{-1}$.\\

As mentioned in Refs.\cite{Huang,Dong}, our model wave function is
obtained by dynamically solving the bound-state RGM equation of the
six quark system in the framework of the extended chiral $SU(3)$
quark model, where the one-gluon-exchange and Goldstone Boson
exchange interactions between quarks are explicitly
considered. Then, by projecting the obtained
wave function onto the inner cluster
wave functions of the $\Delta\Delta$ and CC
channels, the wave function of $d^*$ can be
abbreviated to a form of
\eq \label{wf1} \Psi_{d^*} &=&
[~\phi_{\Delta}(\vec{\xi}_1,\vec{\xi}_2)~\phi_{\Delta}(\vec{\xi}_4,\vec{\xi}_5)~
\chi_{\Delta\Delta}(\vec{R})~\zeta_{\Delta\Delta}~ +~
\phi_{C}(\vec{\xi}_1,\vec{\xi}_2)~\phi_{C}(\vec{\xi}_4,\vec{\xi}_5)~
\chi_{CC}(\vec{R}))~\zeta_{CC}~]_{(SI)=(30)},
\en
where $\phi_{\Delta}$, and $\phi_{C}$ denote the inner cluster wave
functions of $\Delta$ and $C$ (color-octet particle) in the
coordinate space, $\chi_{\Delta\Delta}$ and $\chi_{CC}$ represent
the channel wave functions in the $\Delta\Delta$ and CC
channels  (in the single $\Delta\Delta$ channel case, the $CC$
component is absent), and $\zeta_{\Delta\Delta}$ and $\zeta_{CC}$
stand for the spin-isospin wave functions in the hadronic
degrees of freedom  in the $\Delta\Delta$ and $CC$ channels,
respectively~\cite{Huang}. It should be specially mentioned that in
such a $\ds$ wave function, two channel wave functions are
orthogonal to each other and contain all the totally
anti-symmetrization effects implicitly~\cite{Huang}.\\

In terms of the obtained $\ds$ wave function eq.~(\ref{wf1}) (its
explicit forms have been plotted in Ref.~\cite{Dong}) and, we are
able to calculate the width of the three-body single-pion
decay $d^*\to NN\pi$. The partial width reads
\eq \Gamma_{d^*\to
NN\pi}=\frac{1}{2!}\int d^3p_{1}d^3p_{2}(2\pi)\delta(\Delta E) \Big
| \overline{{\cal M}(\vec{p}_{1},\vec{p}_{2})}\Big |^2,
\en
where $\Big |\overline{{\cal M}(\vec{p}_1,\vec{p}_{2})}\Big |^2$ stands
for the squared transition matrix element with a sum over the
polarizations of the final three body states and an
average of the ones of the initial state $\ds$, the factor of $2!$
is due to the property of the identical
particle in the final $NN$ system, and $\delta(\Delta E)$
denotes the energy conservation with $\Delta
E=M_{d^*}-\omega_{\pi}(k)-E_N(p_1)-E_N(-p_1-k)$ where
$\omega_{\pi}(k)$ and $E_{N}$ represent the energies of the pion and
nucleon, respectively.\\

The transition matrix contains contributions from 12 Feynman diagrams,
where the $\Delta\Delta$ component in the $\ds$ wave function is
responsible for the decay, the pion-exchange is considered for those
sub-leading effects, and the intermediate nucleon state is taken into
account only. Among these diagrams, 6 of them where the outgoing pion
is emitted from $\Delta_2$ are drawn in Fig.1, and they are depicted
according to the time-order perturbation theory.\\

\begin{center}
\begin{figure}[htbp]
{\hskip -6.0cm}
\epsfig{file=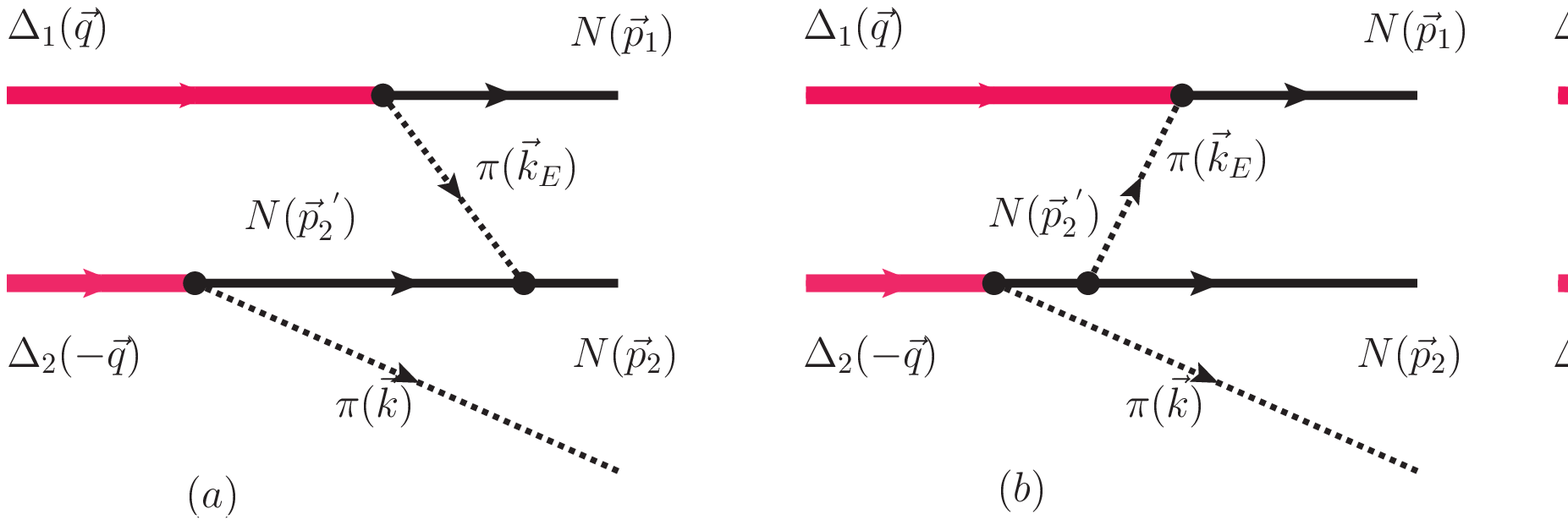,width=10cm, height=3cm}\\
\vspace{0.5cm} {\hskip -6.0cm}
\epsfig{file=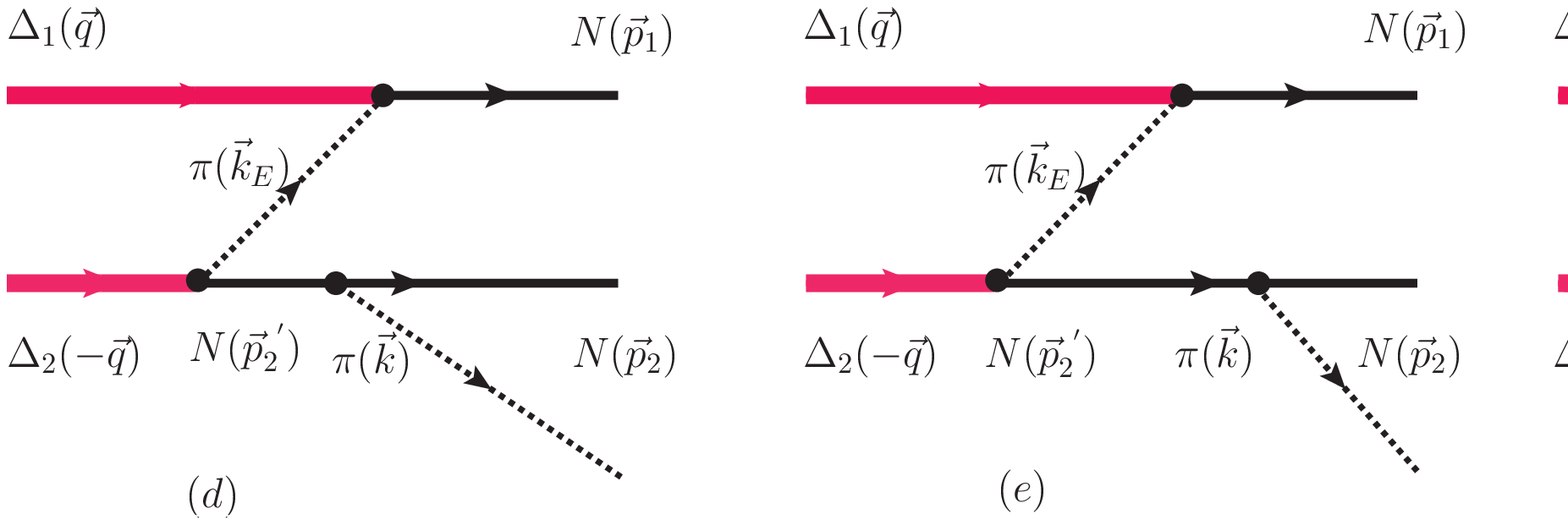,width=10cm,height=3cm}
\caption{Six possible ways to emit pion
only from the $\Delta\Delta$ component of $\ds$ in the
$\ds \to NN \pi$ decay process. The outgoing pion with momenta
$\vec{k}$ is emitted from $\Delta_2$. The other six
sub-diagrams with pion emitted from $\Delta_1$ are similar,
and then are not shown here for reducing the size of the figure.}
\end{figure}
\end{center}
\noindent

Computing transition matrix elements for all the diagrams
in Fig. 1 is straight forward (refer to the discussions in
Ref. ~\cite{Machleidt}). For example, the matrix
element for Fig. 1(a) can be written as
\eq
{\cal M}_{\ds\to NN\pi}^{(a)}&=&\int d^3q
\frac{\Psi_{\ds}(q)}{2\omega_{k_E}\sqrt{2\omega_k}(2\pi)^{6}}
\delta^3\big (p_{N_2'}+p_{N_1'}+k-p_{\Delta_1}-p_{\Delta_2}\big )\nonumber \\
&&\times \tilde{\cal M}_{\pi(k_E)N(p_2')\to N(p_2)}{\cal D}_{af}
\tilde{\cal M}_{\Delta_1\to\pi(k_E)N(p_1)}{\cal D}_{ai} \tilde{\cal
M}_{\Delta_2\to\pi(k)N(p_2')},
\en
where $\Psi_{\ds}$ represents the
$\ds$ wave function in the momentum space which can be obtained by
Fourier transforming the $\ds$ wave functions in the coordinate
space in both the single $\Delta\Delta$ channel and
coupled $\Delta\Delta+CC$ channel approximations (see details in
Refs.~\cite{Dong}) $\tilde{\cal M}_{\pi(k_E)N(p_2')\to N(p_2)}$,
$\tilde{\cal M}_{\Delta_1\to\pi(k_E)N(p_1)}$, and $\tilde{\cal
M}_{\Delta_2\to\pi(k)N(p_2')}$ denote the transitions of
$\pi(k_E)N(p_2')\to N(p_2)$, $\Delta_1\to\pi(k_E)N(p_1)$, and
$\Delta_2\to\pi(k)N(p_2')$, respectively, and the
non-relativistic energy propagators are
\eq
{\cal D}_{af}&=&\frac{1}{M_{\ds}-\omega(\vec{k})-\omega(\vec{k}_E)
-E_N(\vec{p}_1)-E_N(\vec{p}_2^{~'})}\nonumber \\
{\cal D}_{ai}&=&\frac{1}{M_{\ds}-\omega(\vec{k})-E_{\Delta_1}(\vec{q})-E_N(\vec{p}_2^{~'})}.
\en
Then the explicit form of the matrix element (Fig.
1(a)) for the case where the spin of the final two-nucleon
is zero ($s_{12}=0$) can be expressed as
\eq
{\cal
M}_{s_{12}=0}^{(a)}(p_1,k)&=&-\frac{G_{\pi NN}G_{\pi\Delta
N}^2\pi}{(2\pi)^{6}\sqrt{2\omega_k}}\frac{2\sqrt{2}}{27}kY_{1,0}(\hat{k})
\int \frac{d^3q\psi_{d*}(q)k_E^2}{2\omega_{k_E}{\cal D}_{ai}{\cal
D}_{af}}\sum_{m_l}Y_{2m_l}(\hat{k}_E) \times
\sqrt{\frac25}C_{10,2m_l}^{3m_{d*}}.
\en
In the above equation,
$\vec{k}_E=\vec{q}-\vec{p}_1$ stands for the three-momenta of the
exchanged pion, $G_{\pi NN}\cong 4\sqrt{2}G$ and $G_{\pi
N\Delta}\sim \frac{3g_{NN\pi}}{2M_N}$ (with $g_{NN\pi}\sim 13.6$), and
$\vec{k}_E=\vec{q}-\vec{p}_1$.
Moreover, for the case of $s_{12}=1$, we have \eq {\cal
M}_{s_{12}=1}^{(a)}(p_1,k)&=&-\frac{G_{\pi NN}G_{\pi\Delta
N}^2\pi}{(2\pi)^{6}\sqrt{2\omega_k}}\frac{2\sqrt{2}}{27}kY_{1,0}(\hat{k})
\int \frac{d^3q\psi_{d*}(q)k_E^2}{2\omega_{k_E}{\cal D}_{ai}{\cal
D}_{af}}\sum_{m_l}Y_{2m_l}(\hat{k}_E)
\nonumber \\
&\times&  \Big [\sqrt{\frac15}C_{10,1m_{s_{12}}}^{1m_{s_{12}}}C_{1m_{s_{12}},2m_l}^{3m_{d*}}
+\sqrt{\frac23}C_{10,2m_{s_{12}}}^{1m_{s_{12}}}C_{2m_{s_{12}},2m_l}^{3m_{d*}}\Big ].
\en
\\

In general, the final state interaction (FSI) between
two outgoing nucleons should be considered. However, being aware of
the fact that the quantum numbers of $\ds$ are $I(J^P)=0(3^+)$, the
maximal spin of two nucleons is 1, and parity $P$ and total angular
momentum $J$ of the decaying system should be conserved, either the
orbital angular momentum between the outgoing pion and nucleon is at
least equal to 3, or the orbital angular momentum between two
outgoing nucleons at least equals to 2. In the former case, the
decay cross section would be greatly suppressed by the higher
partial wave. And in the latter case, the FSI effect could be
ignored because in the low energy region, the dominant contribution
comes from the S-wave and P-wave, and the contribution from the
higher partial wave can be ignored~\cite{FSI}. Therefore, in this
calculation, we assume the enhancement factor from FSI is close to
1, and consequently would not be considered.\\

\begin{center}
{\bf Table 1, The calculated decay width of the
$\ds\to pn\pi^0$ process and the widths
contributed individually from the (a)-, (b)-, (c)-, (d)-, (e)-, and
(f)-type diagrams (in units of MeV).}\\
\vspace{0.5cm}
\begin{tabular}{|c||c||c|c|c|c|c|c|c|}\hline
Case &Total width&(a) &(b) &(c) &(d) &(e) &(f) &sum of (a)-(f)\\
\hline One ch. ($\Delta\Delta~only$)   &2.276 &~~0.550~~
&~~0.306~~&~~0.267~~ &~~0.0963~~ &~~0.209~~ &~~0.233~~ &1.661
\\ \hline Two chs. ($\Delta\Delta$+CC)   &0.670 &0.154 &0.0884 &0.0789 &0.0279 &0.0687 &0.0847
&0.503\\ \hline
\end{tabular}
\end{center}

The numerical results for the decay width in
the $\ds\to pn\pi^0$ process and the widths
contributed individually from the (a)-, (b)-, (c)-,
(d)-, (e)-, and (f)-type diagrams are tabulated in Table
1, respectively. Contributions from the diagrams where the
outgoing pion being emitted from $\Delta_1$ are also taken
into account. The total decay width of the $\ds
\to NN\pi$ process is the sum of the contributions from
all types of diagrams coherently. From table 1, one sees that the
ratio of the decay width with coherent sum to
that with incoherent sum is about 1.37 in the one channel
($\Delta\Delta$ channel only) case and about 1.33 in the two channel
(coupled $\Delta\Delta$+CC channels) case, respectively, which shows
a sizeable coherent effect. The most important issue from
this table is that the decay width of $\ds$ to $pn\pi^0$ is
smaller than 3 MeV in the one channel case and about 670 KeV in the
two channel case.  Apparently, the width in the
two channel case is much smaller than that in the one channel
case. Since  in the two channel case, the
contribution comes from the  $\Delta\Delta$
component of $\ds$ only, which is about 31.5\% of the
whole $\ds$ wave function, but in the one channel
calculation, the contribution comes from the whole $\Delta\Delta$
wave function, therefore, this outcome is understandable.
The second observation is that although the framework and
method in this calculation are  the same as that used in
the $\ds$ width  calculations in the
double-pion decay processes before, the obtained decay
widths for the single-pion decay mode are remarkably
smaller than those given in our previous
calculations~\cite{Dong,Dong1} and the experimental data
for the double-pion decay mode. This is
because that in the single pion decay process,
the leading non-vanishing contribution comes from the
sub-leading diagrams shown in Fig. 1, where three vertices
exist, whereas in the double-pion decay process, the
leading non-vanishing contribution comes from the
two vertices diagrams shown in Refs.~\cite{Dong,Dong1}.
Clearly, the obtained very small decay width for the
single pion decay mode is consistent with the current experimental
status that no $\ds \to NN \pi$ process has been found
in the present data set.\\

Some approximations in calculation should be further
discussed. In the sub-leading diagrams shown in Fig. 1,
the Goldstone-Boson exchange between two nucleons must be
introduced to convert two $\Delta$s to two nucleons.
From PDG~\cite{PDG}, one finds that the largest decay
mode for $\Delta(1232)3/2^+$ is $N\pi$ with a branching ratio of
about 100\%. Therefore, in the realistic calculation, considering
the pion-exchange only would not miss the major feature and
make the result meaningless. Due to the larger mass of nucleon
excitations, we do not take them as  the intermediate nucleon
state. We also ignore the contribution from the diagrams
where the intermediate $\Delta$ state exist, because the quark model
calculation tells us that the $\Delta\Delta\pi$ coupling
$f_{\Delta\Delta\pi}$ is much smaller than the $NN\pi$ coupling
$f_{NN\pi}$~\cite{AREN}. Again, we would
specially emphasize that the contributions from the large
CC component in the $\ds$ wave function could be
ignored. The reason is the following. In the previous
paragraph, we have mentioned the extracted channel wave functions
for various channels have already absorbed the effect of the totally
anti-symmetrization, and the channel wave functions for the
$\Delta\Delta$ and CC channels are orthogonal to each other. So in
our decay calculation, the inter-cluster quark exchange should not
be considered anymore. In the $(SI=30)$ case, where $S$ and $I$
denote the spin and isospin of the system, respectively, converting
CC to $\Delta\Delta$ requires an exchange of a colored object,
namely a gluon. The calculation shows that without quark exchange,
the matrix elements of the one-gluon-exchange interaction (OGE)
between $\Delta\Delta$ and CC are zero, namely CC cannot be
converted to $\Delta\Delta$. If one would convert CC$(SI=30)$ to
$\Delta\Delta(S\neq 3)$ and CC to NN, because the spin of the system
is $S=3$ and the parity is positive, it needs at least D-wave between
$\Delta$ and $\Delta$ with ($S\neq 3$) and between N and N, respectively.
Then, the required tensor force in OGE, which is a higher order term,
would make these conversions suppressed greatly.  In short, we can ignore
the contribution from the CC component in $\ds$ in the $\ds$ decay
calculation. \\

To summarize, we proceed a calculation for the single-pion decay
mode of $\ds$ with the help of our wave function obtained
in the chiral constituent quark model calculation.
It shows in our calculations~\cite{Huang,Dong,Dong1} that the CC
component has a large fraction of about 2/3 in the wave
function of $\ds$, and the rest part, the $\Delta\Delta$
component, is responsible for its widths in the decays of
$\ds\to d\pi\pi$ and $\ds\to NN\pi\pi$ as well as $\ds\to NN\pi$. As a result,
the obtained partial widths for
the double-pion decay modes of $\ds$ are in good  agreement
with the experimental measurement. Moreover, the
obtained width for the single-pion decay model in this
calculation is much smaller than those in the double-pion decay modes.
If we assume that the total width of the $\ds$ is about 75~MeV, the
predicted branching ratio of the single-pion decay mode $\ds \to
pn\pi^0$ is about 3.0\% in the one-channel case and  0.9\% in the
coupled-channel case. It should be emphasized that the obtained
single-pion decay width is much smaller than the
widths of the double-pion decay modes.
This result agrees with the present experiment status that
such a single-pion decay mode has not been found in that data
analysis. It is quite different from the width
reported by the investigation with the $\Delta N \pi$ assumption for
the structure of $\ds$, where the predicted width for
the single pion decay mode would be large enough to
be observed in the experimental measurement. It is
expected that an intensive data analysis of the $\ds \to
pn\pi^0$ channel would judge different explanations
for the nature of $\ds$.\\

\begin{acknowledgments}

We would like to thank Heinz Clement, Qiang Zhao, Bing-Song Zou, Xu
Cao, and Qi-Fang L\"u for their useful and constructive discussions.
F. Huang is grateful for the support of the Youth Innovation Promotion
Association of CAS under the grant No. 2015358. This work is
supported by the National Natural Sciences Foundations of China under the
grant Nos. 11475192, 11475181, 11521505, 11565007, and
11635009, and by the fund provided to the Sino-German CRC 110 ``Symmetries and
the Emergence of Structure in QCD" project by the DFG, and the IHEP Innovation
Fund under the No. Y4545190Y2.
\end{acknowledgments}

\end{document}